\documentclass[preprint,prl,amsmath]{revtex4}

\usepackage{graphicx}
\usepackage{amssymb,amsmath, wasysym}
\usepackage{natbib}
\bibpunct{}{}{,}{s}{}{} % superscript citation

\usepackage[english]{babel}

\usepackage[euler]{textgreek}

\newcommand{\Fourier}[1]{\mathcal{F}\left\{#1\right\}}

\begin{document}

\title{Exploiting speckle correlations to improve the resolution of \\ wide-field fluorescence microscopy}

\author{Hasan Y{\i}lmaz$^{1}$, Elbert G. van Putten$^{1,2}$, Jacopo Bertolotti$^{1,3}$, Ad Lagendijk$^{1}$, Willem L. Vos$^{1}$, and Allard P. Mosk$^{1}$}
\affiliation{$^{1}$Complex Photonic Systems (COPS), MESA+ Institute for Nanotechnology,\\
University of Twente, P.O. Box 217, 7500 AE Enschede, The Netherlands\\
$^{2}$Present address: Philips Research Laboratories, 5656 AE Eindhoven, The Netherlands\\
$^{3}$Present address: Physics and Astronomy Department, University of Exeter,\\
Stocker Road, Exeter EX4 4QL, United Kingdom\\
}

\maketitle

{\bfseries
Fluorescence microscopy is indispensable in nanoscience and biological sciences. The versatility of labeling target structures with fluorescent dyes permits to visualize structure and function at a subcellular resolution with a wide field of view. Due to the diffraction limit, conventional optical microscopes are limited to resolving structures larger than 200 nm. The resolution can be enhanced by near-field and far-field super-resolution microscopy methods. Near-field methods typically have a limited field of view and far-field methods are limited by the involved conventional optics. Here, we introduce a combined high-resolution and wide-field fluorescence microscopy method that improves the resolution of a conventional optical microscope by exploiting correlations in speckle illumination through a randomly scattering high-index medium: Speckle correlation resolution enhancement (SCORE). As a test, we collect two-dimensional fluorescence images of 100-nm diameter dye-doped nanospheres. We demonstrate a deconvolved resolution of 130 nm with a field of view of $\mathbf{10\times10}$ \textmu$\mathbf{m^2}$.}

A conventional optical microscope produces images with a resolution determined by the numerical aperture (NA) of the imaging lens. The NA of an imaging lens is defined by the highest wave vector that is accessible in the transversal direction.  Many methods have been introduced that enable optical resolution beyond the resolution limit of a conventional optical microscope, by exploiting evanescent waves with near-field scanning optical microscopy (NSOM) \cite{NSOM_1986}, by exploiting moir\'{e} fringes as in structured-illumination microscopy (SIM) \cite{1999_proc_SIM, SIM_2000}, by exploiting nonlinear optical phenomena as in saturated structured-illumination microscopy (SSIM) \cite{2005_NSIM} or stimulation emission depletion (STED) microscopy \cite{Hell1994aa, Hell2007}, by exploiting specific photophysical properties of dyes as in stochastic optical reconstruction microscopy (STORM) \cite{Rust2006aa} or photoactivated localization microscopy (PALM) \cite{Betzig2006}. Nevertheless, NSOM has a field of view limited by the scan range of the probe, moreover a scanning probe usually greatly affects the measurement itself. SIM provides a resolution that is potentially two times higher than a conventional optical microscope \cite{1999_proc_SIM, SIM_2000}. SIM requires a precise knowledge of the illuminating field on the structure of interest. Recently the statistical properties of unknown speckle patterns were exploited to relax the requirements on precise knowledge of the illumination field \cite{2012_bSIM_sentenac}. Since STED requires intense laser pulses, it is a question how to use it for delicate samples with a low damage threshold. STED, STORM and PALM require dyes with specific photophysical properties. Although abovementioned far-field microscopy methods realize an optical resolution beyond the diffraction limit, the resolution remains strongly dependent on the NA of their conventional optics. It has been shown that a scattering medium enhances spatial resolution both for acoustic waves \cite{Derode2003_PRL, Lerosey2007aa} and light waves \cite{Vellekoop2010aa}, and turns a high-index substrate into a high-NA solid immersion lens by breaking the translational invariance on the interface of the substrate \cite{vanPutten2011_PRL}. Coherent light illumination on such a scattering medium generates a speckle pattern of apparently randomly distributed bright and dark regions behind the scattering medium. The concept of exploiting correlations between such speckle patterns has started a new class of optical microscopy \cite{Freund_1990}. Within a speckle pattern there is a correlation effect called the optical memory effect \cite{Freund1988aa, Feng1988aa}, that has recently been exploited for optical imaging through scattering media \cite{Hsieh2010aa, Vellekoop2010ab, vanPutten2011_PRL, silbelberg_2012, 2013_zhou_opt.express, bertolotti_nature_2012, psaltis_seethrough_2014, katz_seethrough_2014}. A tilt of the incident beam within an angle of $\Delta\theta < \lambda/2\pi d$ ($\lambda$ is the wavelength of light and $d$ the thickness of the scattering medium) results in a translation of the speckle pattern behind the medium without a significant change in the pattern. The optical memory effect has been employed to obtain optical images of microscopic objects hidden by a scattering medium \cite{bertolotti_nature_2012, psaltis_seethrough_2014, katz_seethrough_2014}. Previously, a high-resolution gallium phosphide (GaP) scattering lens has been used to image gold nanoparticles with elastically scattered coherent visible light \cite{vanPutten2011_PRL}. However, the available field of view with speckle correlations is limited to 2$\times$2 \textmu$\mathrm{m^2}$ due to the finite range of the optical memory effect, and the high-resolution scattering lens has not been applied to incoherent imaging modalities such as fluorescence microscopy.

\begin{figure}[htb]
\begin{center}
\includegraphics[width=8.8 cm]{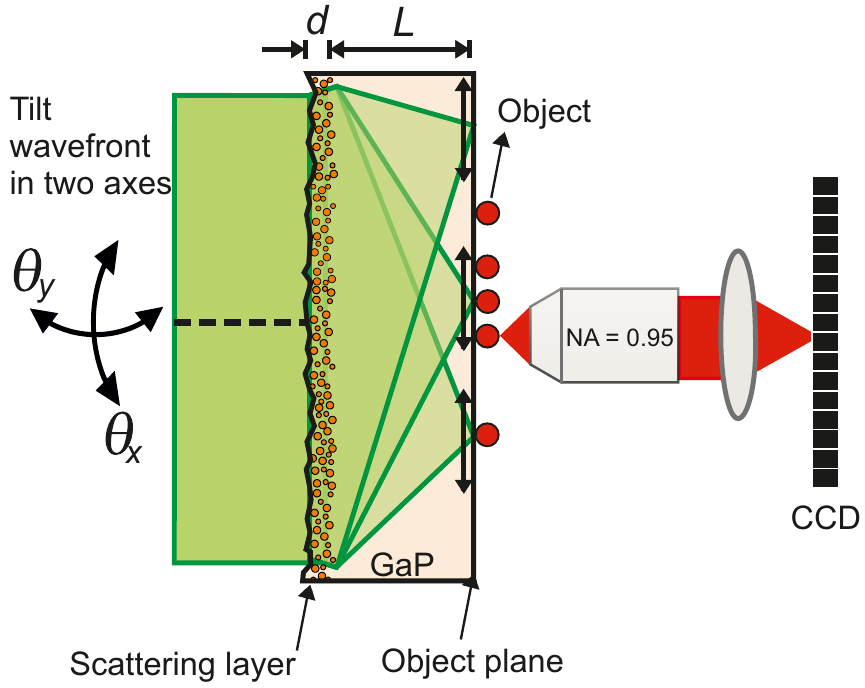}
\caption{The concept of the wide-field, high-resolution speckle scanning microscopy. A beam of coherent light illuminates a scattering layer on a gallium phosphide (GaP) substrate. The scattered light generates a speckle pattern that enables high-resolution imaging on the object plane. The incident beam is tilted by angles $\theta_x$ and $\theta_y$, and as a result the speckle pattern is scanned across the sample. Fluorescent nanospheres on the object plane are imaged on a CCD camera. With parallel speckle detection, the field of view is wider than the single speckle-scan range. (Acronyms: $d$, the thickness of the scattering layer; $L$, the thickness of the substrate).}
  \label{fig:exp_concept}
    \end{center}
\end{figure}

Here we introduce and demonstrate a fluorescence microscopy method that combines the high-resolution of speckle scanning microscopy with a wide field of view of parallel speckle-scan detection. Fig. \ref{fig:exp_concept} shows the concept of our method. The main element of the experiment is a scattering lens, consisting of a GaP substrate with a thickness of $L = 400$ \textmu m and a refractive index of $n = 3.42$ of which one surface has a scattering layer with a thickness of $d = 2$ \textmu m and one surface is polished \cite{vanPutten2011_PRL}. A beam of coherent light with a diameter of 0.8 mm and a wavelength of $\lambda_{ill} =$ 561 nm is incident onto the scattering surface of the substrate. The scattering layer generates a speckled intensity pattern $S(\Delta x,\Delta y)$ that illuminates a fluorescent object $O(\Delta x,\Delta y)$ on the object plane that is on the polished surface of the GaP substrate. The fluorescence intensity distribution on the object plane is imaged on a camera with a resolution of $R = \lambda_{flu} /(2\text{NA})$, with $\lambda_{flu}$ = 612 nm and NA = 0.95. We raster scan the speckle pattern on the object plane by tilting the incident beam by angles $\delta\theta_x$ and $\delta\theta_y$, within the angular range of the optical memory effect ($\Delta\theta \simeq 1\,^{\circ}{\rm}$) that leads to a speckle-scan range of 2 \textmu m on the object plane. We record fluorescence images at every $\delta\theta_x$, $\delta\theta_y$ for a range of angles of incidence, resulting in perpendicular speckle pattern displacements $\delta x\simeq\delta\theta_x\ L/n$ and $\delta y\simeq\delta\theta_y\ L/n$ in the object plane. This procedure results in a four-dimensional data-cube $I(\Delta x,\Delta y,\delta x,\delta y)$, which has all information to reconstruct a wide-field image that has a resolution of the average speckle grain size \cite{bertolotti_nature_2012}.

\section*{High-resolution information retrieval}

\medskip

\begin{figure}
\begin{center}
\includegraphics[width=15 cm]{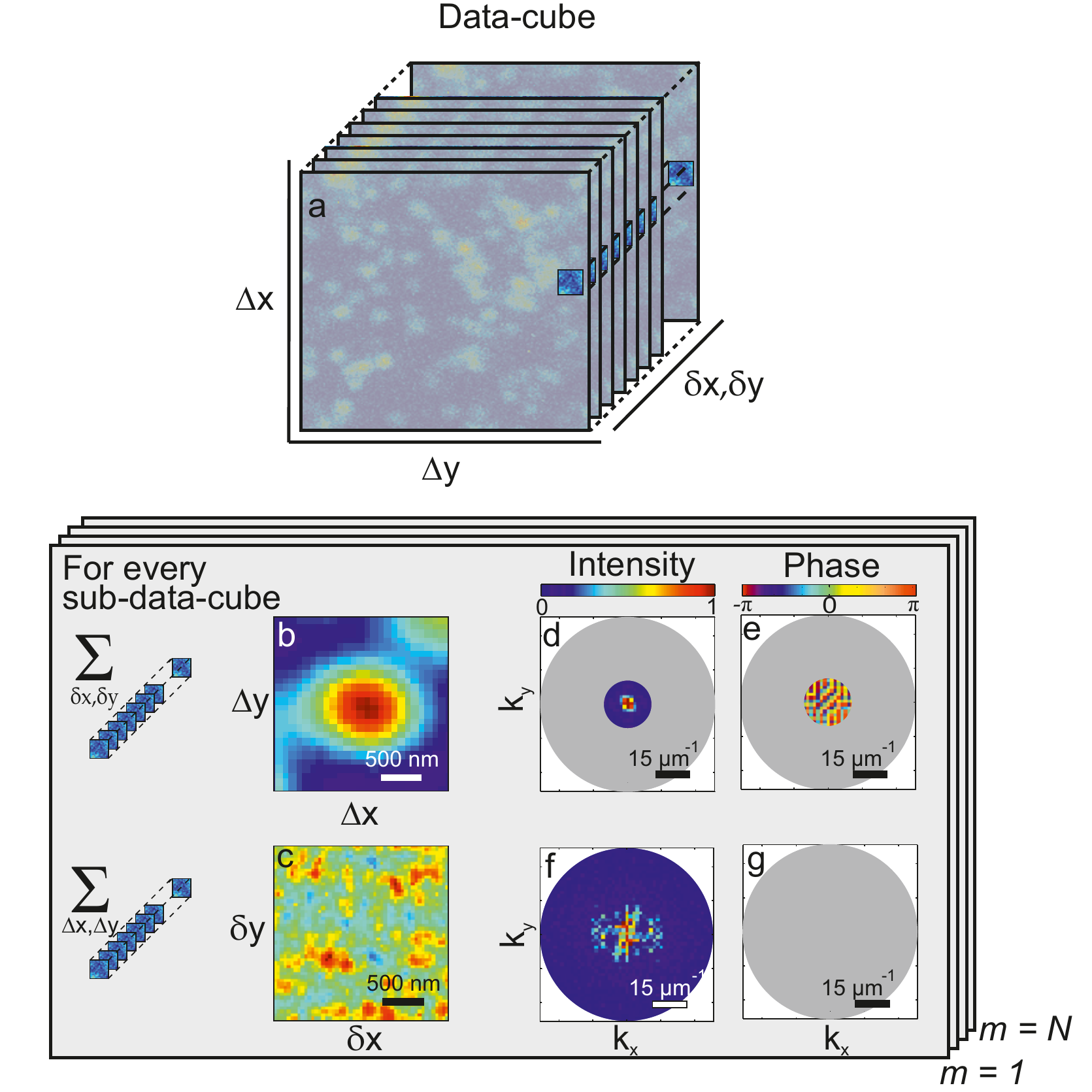}
\caption{The data analysis procedure on a single sub-data-cube. $\boldsymbol{\mathrm{a}}$: Data-cube $I(\Delta x,\Delta y,\delta x,\delta y)$. $\boldsymbol{\mathrm{b}}$: The obtained standard resolution sub-image $J_m(\Delta x, \Delta y)$ by summing the sub-data-cube shown by the square non-greyed out area over $\delta x$ and $\delta y$. $\boldsymbol{\mathrm{c}}$: The obtained speckle-scan matrix $K_m(\delta x, \delta y)$ by summing the sub-data-cube shown by the square non-greyed out area over $x$ and $y$. \textbf{d}: The intensity of the Fourier components of $J_m(\Delta x, \Delta y)$. $\boldsymbol{\mathrm{e}}$: The phase of the Fourier components of $J_m(\Delta x, \Delta y)$. \textbf{f}: The intensity of the Fourier components of $K_m(\delta x, \delta y)$. \textbf{g}: The phase of the Fourier components of $K_m(\delta x, \delta y)$.}
\label{fig:procedure}
  \end{center}
\end{figure}

In Fig. \ref{fig:procedure} we show the data analysis procedure. We divide the data-cube (Fig. \ref{fig:procedure}a) into $N$ sub-data-cubes by applying $N$ square window functions of $W_m(x,y)$ that each have a width and a height equal to half of the speckle-scan range (1 \textmu m) that each can be processed in parallel. We construct a standard resolution sub-image $J_m(\Delta x,\Delta y)$ (Fig. \ref{fig:procedure}b) and a speckle-scan matrix $K_m(\delta x, \delta y)$ (Fig. \ref{fig:procedure}c) from the corresponding sub-data-cube as follows: We sum our sub-data-cube over $\delta x$ and $\delta y$, and obtain the standard resolution sub-image $J_m(\Delta x, \Delta y)$. In our approach, it is useful to represent $J_m(\Delta x,\Delta y)$ in the Fourier domain, where its spatial information is given by the intensity and the phase of the Fourier components (Fig. \ref{fig:procedure}d,e). To obtain the speckle-scan matrix $K_m(\delta x,\delta y)$, we calculate the following summation
\begin{align}
K_m(\delta x,\delta y)&= \sum\limits_{\Delta x, \Delta y}I(\Delta x,\Delta y,\delta x,\delta y)W_m(\Delta x,\Delta y)
\nonumber\\
&= \sum\limits_{\Delta x, \Delta y}O(\Delta x,\Delta y)S(\Delta x - \delta x,\Delta y - \delta y)
\nonumber\\
W_m(\Delta x,\Delta y)
\nonumber\\
&=[(O\cdot W_m)\ast S](\delta x, \delta y),
\label{eq:convolution}
\end{align}
where the symbol $\ast$ denotes a convolution product and where in the last step we assumed that the scan range stays within the optical memory effect range. In Figs. \ref{fig:procedure}f and \ref{fig:procedure}g we represent the speckle-scan matrix $K_m(\delta x, \delta y)$ in the Fourier domain. We obtain the intensity of the high-frequency Fourier components of the object from its speckle-scan matrix as follows:
\begin{align}
\vert \Fourier{K_m}\vert &=\vert \Fourier{O\cdot W_m}\vert\cdot \vert \Fourier{S}\vert
\nonumber\\
&= C\vert\Fourier{O\cdot W_m}\vert
\label{eq:Fourier_modulus}
\end{align}
where $C$ is the autocorrelation of the amplitude transfer function of our scattering lens, and $\Fourier{}$ denotes a Fourier transform. Here we use the approximation that within the NA of the GaP scattering lens, the absolute value of the spatial spectrum of the field is constant for a fully developed speckle pattern \cite{Goodman2000}. Equation \ref{eq:Fourier_modulus} shows that the intensity of the high-frequency Fourier components of the object is retained behind the scattering layer (Fig. \ref{fig:procedure}f). The phase information of the object's Fourier components is lost due to the random and unknown phase of the speckle pattern (Fig. \ref{fig:procedure}g). Fortunately, it is often possible to infer the lost phase information using an iterative phase retrieval algorithm \cite{Fienup1978_OL, Fienup1982_AO, Millane1990_JOSAA, segev_2012_natmater}. In essence, our approach relies on reducing the light scattering problem into a phase retrieval problem.

\section*{Image reconstruction}

\medskip

\begin{figure}
\begin{center}
\includegraphics[width=8.8 cm]{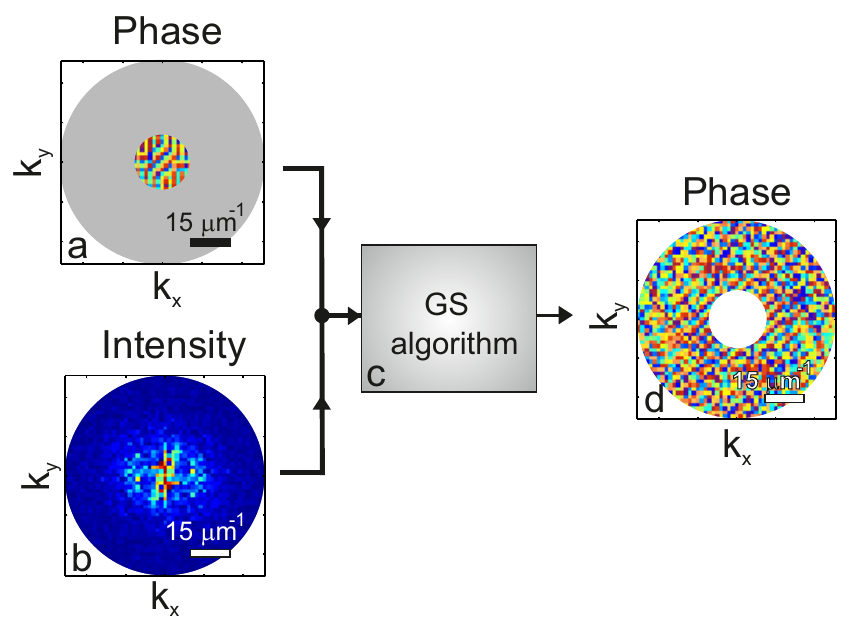}
\caption{Phase retrieval in the Fourier domain. $\boldsymbol{\mathrm{a}}$: The phase of the Fourier components of the object. $\boldsymbol{\mathrm{b}}$: The intensity of the Fourier components of the object. $\boldsymbol{\mathrm{c}}$: The Gerchberg-Saxton-type algorithm. $\boldsymbol{\mathrm{d}}$: The retrieved phase of high-frequency Fourier components of the object. (The phase data comes from Fig. \ref{fig:procedure}e and the intensity data comes from Fig. \ref{fig:procedure}f. Colourbars are as in Fig. \ref{fig:procedure}).}
\label{fig:phase_retrieval}
  \end{center}
\end{figure}

We have developed a new Gerchberg-Saxton-type algorithm that uniquely retrieves the high-frequency phase information of the Fourier components of our object using the low-frequency phase information of the Fourier components of the object as constraint. In general, a Gerchberg-Saxton-type algorithm retrieves the phase of the Fourier components of an image from its intensity of the Fourier components with some constraints on the image such as consisting of real and positive numbers. In a Gerchberg-Saxton-type algorithm, using only the intensity of the Fourier components gives ambiguities in the solution \cite{Millane1990_JOSAA, segev_phase_retrieval_2014}. These ambiguities are flips or translations of the reconstructed intensity object. In our Gerchberg-Saxton-type algorithm, we use a standard resolution image of our object to use its phase of the low-frequency Fourier components as additional information to obtain a unique solution. We use constraints both in the object domain and in the Fourier domain. In the object domain we use the information that the measured intensity of our fluorescent object is real and positive. In the Fourier domain, we use the phase of the low-frequency Fourier components. Combining these two types of information the algorithm converges to a unique solution which gives us the shape, the position and the orientation of the object. This is a major improvement over previous approaches \cite{bertolotti_nature_2012, psaltis_seethrough_2014, katz_seethrough_2014} that do not provide position and orientation information.

In Fig. \ref{fig:phase_retrieval} the phase retrieval procedure of high-frequency Fourier components is shown for a single sub-data-cube. First, we Fourier transform both a standard resolution sub-image, $J_m(\Delta x, \Delta y)$ and the corresponding speckle-scan matrix, $K_m(\delta x, \delta y)$. We discard the intensity of the Fourier components of $J_m(\Delta x, \Delta y)$ and the phase of the Fourier components of $K_m(\delta x, \delta y)$. We input the phase information of low-frequency Fourier components of $J_m(\Delta x, \Delta y)$ and the intensity information of high-frequency Fourier components of $K_m(\delta x, \delta y)$ into our Gerchberg-Saxton-type algorithm. The algorithm outputs the phase information of high-frequency Fourier components. Finally, we combine and inverse Fourier transform all available phase and intensity information of the Fourier components to obtain the high-resolution sub-image.

To acquire a wide-field image, we apply our phase retrieval procedure (see Fig. \ref{fig:phase_retrieval}) to every sub-data-cube (see Fig. \ref{fig:procedure}) in parallel. Each reconstructed overlapping high-resolution sub-image is windowed by a smooth window function to minimize edge effects. We tile the reconstructed high-resolution sub-images to yield a wide-field image of the complete object. The field of view of the reconstructed image is wider than the speckle-scan range and spans the field of view of the detection optics.

\section*{Discussion}

\medskip

\begin{figure}
\begin{center}
\includegraphics[width=15 cm]{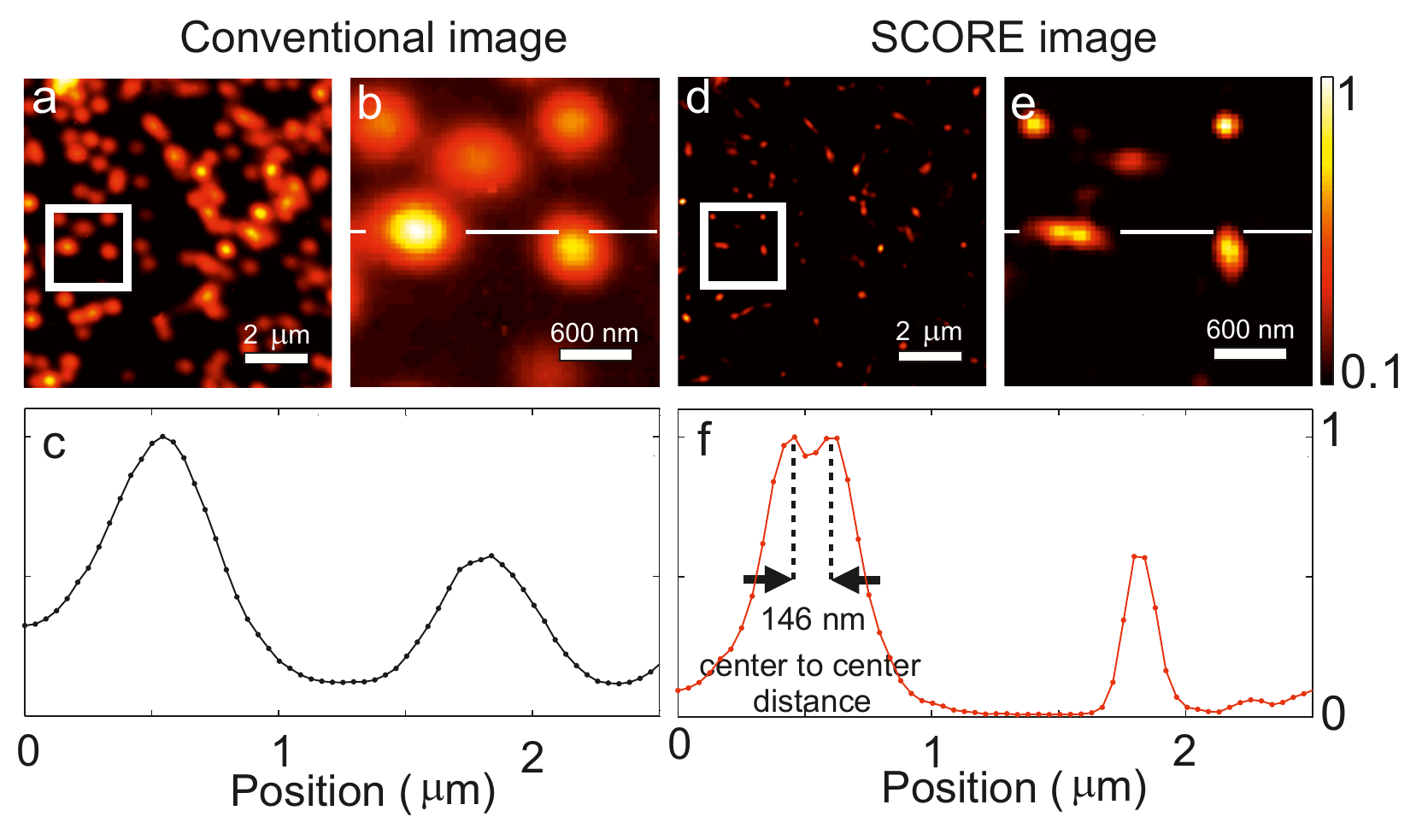}
\caption{Wide-field images of fluorescent nanospheres with diameter of 100 nm. \textbf{a}: The wide-field image by conventional microscopy. \textbf{b}: A zoomed image of \textbf{a}. \textbf{c}: A cross section of \textbf{b} represented by the white line. \textbf{d}: The wide-field image by SCORE microscopy. \textbf{e}: A zoomed image of \textbf{d}. \textbf{f}: A cross section of \textbf{e} represented by the white line. In \textbf{c}, a single nanosphere is apparent while in \textbf{f} two smaller nanospheres are apparent with a center to center distance of 146 nm from each other.}
\label{fig:result}
  \end{center}
\end{figure}

To experimentally test our new imaging method we use a collection of fluorescent nanospheres with a diameter of 100 nm as test objects. Fig. \ref{fig:result}a shows an image of a collection of many fluorescent nanospheres taken with conventional high-NA microscopy in a field of view of $10\times10$ \textmu m$^2$. The zoom-in in Fig. \ref{fig:result}b reveals five separate nanospheres. Fig. \ref{fig:result}c shows a cross-section of two nanospheres from Fig. \ref{fig:result}b that have a full-width-half-maximum of about 450 nm. We now turn to the high-resolution SCORE results. Fig. \ref{fig:result}d shows the same area as in Fig. \ref{fig:result}a. In Fig. \ref{fig:result}d the nanospheres are sharper compared to the image in Fig. \ref{fig:result}a. The zoom-in in Fig. \ref{fig:result}e shows the same area as in Fig. \ref{fig:result}b: We see that the nanospheres are much sharper compared to Fig. \ref{fig:result}b and we see six separate nanospheres, whereas less nanospheres were discernible in Fig. \ref{fig:result}b. Notably at the left center two nanospheres are distinguished that were observed as one blob on Fig. \ref{fig:result}b. Fig. \ref{fig:result}f shows a cross-section of three nanospheres from Fig. \ref{fig:result}e. A clear demonstration of the enhanced resolution is given in Fig. \ref{fig:result}f where we clearly resolve two nanospheres with a center to center distance of 146 nm, and an edge to edge distance of 46 nm. A numerical deconvolution of the image of a single nanosphere with the known shape of the object reveals that we have a resolution of 130 nm according to the Sparrow's criterion. The deconvolved full-width-half-maximum of our point spread function is 140 nm, which is slighty larger than the full-width-half-maximum of $r = 116$ nm expected for the given illumination beam width. The difference between the expected and the demonstrated resolutions may be due to sample drifts during the experiment and pointing noise of the laser. Our results demonstrate that regardless of the range of the optical memory effect, speckle correlations enhance the resolution of an optical microscope without any restriction on its field of view.

In summary, we experimentally demonstrate a new method to obtain high-resolution and wide-field fluorescence images. In combination with a gallium phosphide scattering lens, speckle correlation resolution enhancement (SCORE) has the ability to acquire very high-resolution images with a field of view that is much wider than the speckle-scan range. SCORE is thus excellently suited to be used for imaging of two-dimensional slice of an object as large as a few hundred micrometers with subcellular resolution. Characterization of the scattering medium by methods such as wavefront shaping \cite{vanPutten2011_PRL}, digital optical phase conjugation \cite{Hsieh2010aa} or transmission matrix measurement \cite{Popoff2010ab, Choi2011_PRL} is not needed. 

The resolution of our current proof of principle experimet is limited by signal to noise and stage drift. A higher illumination power, a wider beam, and a shorter excitation wavelength can be used to approach the resolution limit of $\lambda_{ill}/2n = 80$ nm in GaP where $n$ = 3.45 for $\lambda_{ill}$ = 550 nm. Without any additional hardware, resolution of SCORE can be improved up to $(2n/\lambda_{ill}+2NA/\lambda_{flu})^{-1} = 64$ nm by using the resolution information of the conventional microscope objective in detection as in SIM \cite{1999_proc_SIM, SIM_2000, 2005_NSIM, 2012_bSIM_sentenac}.

\section*{Methods}

\medskip

\textbf{Parallel detection:} Speckle-scan matrices contain high-resolution information of imaging object. In order to measure a speckle-scan matrix $K_m(x,y,\delta x,\delta y)$, the speckle pattern has to stay correlated over the resolution $R = \lambda_{flu}/2\text{NA}$. This constraint is met when $R < \pi nL/\lambda_{ill} d$ where $n$ is the refractive index of the GaP substrate, $L$ the thickness of the GaP substrate, $\lambda_{ill}$ the wavelength of the incident light, and $d$ the thickness of the GaP porous layer. In our GaP substrate $\pi nL/\lambda_{ill} d$ is in the order of 2 \textmu m. Our detection optics has a resolution ($R$ = 322 nm) that is high enough to fulfill this condition. The average speckle grain size of a GaP scattering lens is $r = \lambda/[2n\text{sin}(\text{tan}^\text{-1}(W/2L))]$, where $W$ is the beam width. In our case, an average speckle grain size is $r = 116$ nm. We scan the speckle pattern with steps of 40 nm over a range of 2 \textmu m in two dimensions, requiring $N$ = 2500 measurements. For each measurement the full camera image is stored, which allows us to retrieve the object at any position of the captured field of view.

%\bibliographystyle{naturemag}
%\bibliography{ref_score}

%\medskip
%\textbf{Supplementary Information} is linked to the online version of this paper at www.nature.com/nphoton

\medskip

\section*{Acknowledgements}
The authors thank Duygu Akbulut, Femi Ojambati, Henri Thyrrestrup, Pepijn Pinkse, and Sebastianus Goorden for discussions. This work is part of the research program of "Stichting voor Fundamenteel Onderzoek der Materie (FOM)," which is financially supported by "Nederlandse Organisatie voor Wetenschappelijk Onderzoek (NWO). We thank STW, and ERC (279248) for support."

\medskip

\section*{Author Contributions}
All authors take full responsibility for the content of the paper.

\medskip

\section*{Additional Information}
Correspondence and requests for materials should be addressed to HY (h.yilmaz@utwente.nl).

\medskip

\section*{Competing financial interests}
The authors declare no competing financial interests.

\end{document}